\documentclass[twocolumn,showpacs,aps,prd,floatfix,preprintnumbers]{revtex4}
\usepackage{graphicx} 
\usepackage{dcolumn}
\usepackage{epsfig}    
\usepackage{amsmath}
\input babarsym   

\newcommand{\etactwo}{\ensuremath{\eta_{c}(2S)}\xspace}

\renewcommand{\mm}{\ensuremath{M^2_{\mathrm{rec}}}\xspace}

\newcommand{\kketa}{\ensuremath{\kp\km \eta}\xspace}
\newcommand{\kkpiz}{\ensuremath{\kp\km \piz}\xspace}

\newcommand{\etaceta}{\ensuremath{\etac \to \kp\km \eta}\xspace}
\newcommand{\etacpiz}{\ensuremath{\etac \to \kp\km \piz}\xspace}
\newcommand{\etactwoeta}{\ensuremath{\etactwo \to \kp\km \eta}\xspace}
\newcommand{\etactwopiz}{\ensuremath{\etactwo \to \kp\km \piz}\xspace}
\newcommand{\etapipipi}{\ensuremath{\eta \to \pip \pim \piz}\xspace}
\newcommand{\pipipi}{\ensuremath{\pip \pim \piz}\xspace}
\newcommand{\chictwotopiz}{\ensuremath{\chi_{c2} \to \kp\km \piz}\xspace}
\newcommand{\chictwotoeta}{\ensuremath{\chi_{c2} \to \kp\km \eta}\xspace}
\def\Kstarz   {\ensuremath{K^*_0(1430)}\xspace}

\renewcommand{\gg}{\ensuremath{\gamma\gamma}}
\def\calR         {{\ensuremath{\cal R}\xspace}}

\newcommand{\BaBarPubYear}    {13}
\newcommand{\BaBarPubNumber}  {021}
\newcommand{\SLACPubNumber}   {15928}

\begin{document}

\begin{flushleft}
\babar-PUB-\BaBarPubYear/\BaBarPubNumber \\
SLAC-PUB-\SLACPubNumber
\end{flushleft}

\title{
 \large \bf\boldmath Dalitz plot analysis of \etaceta\ and \etacpiz\ in
 two-photon interactions
}

%
\author{J.~P.~Lees}
\author{V.~Poireau}
\author{V.~Tisserand}
\affiliation{Laboratoire d'Annecy-le-Vieux de Physique des Particules (LAPP), Universit\'e de Savoie, CNRS/IN2P3,  F-74941 Annecy-Le-Vieux, France}
\author{E.~Grauges}
\affiliation{Universitat de Barcelona, Facultat de Fisica, Departament ECM, E-08028 Barcelona, Spain }
\author{A.~Palano$^{ab}$ }
\affiliation{INFN Sezione di Bari$^{a}$; Dipartimento di Fisica, Universit\`a di Bari$^{b}$, I-70126 Bari, Italy }
\author{G.~Eigen}
\author{B.~Stugu}
\affiliation{University of Bergen, Institute of Physics, N-5007 Bergen, Norway }
\author{D.~N.~Brown}
\author{L.~T.~Kerth}
\author{Yu.~G.~Kolomensky}
\author{M.~J.~Lee}
\author{G.~Lynch}
\affiliation{Lawrence Berkeley National Laboratory and University of California, Berkeley, California 94720, USA }
% ...
\collaboration{The \babar\ Collaboration}
\noaffiliation

\begin{abstract}
We study the processes $\gg\to \kketa$ and $\gg\to \kkpiz$ using a data sample
of 519~\invfb\ recorded with the \babar\ detector operating at the SLAC PEP-II
asymmetric-energy \epem\ collider at center-of-mass energies at and near the
$\Upsilon(nS)$ ($n = 2,3,4$) resonances.
We observe \etacpiz\ and \etaceta\ decays, measure their relative branching fraction, and perform a Dalitz plot 
analysis for each decay. We observe the  $\Kstarz \to K \eta$ decay and
measure its branching fraction relative to the $K \pi$ decay mode to be 
$\calR(K^*_0(1430)) = \frac{\BR(K^*_0(1430) \to K \eta)}{\BR(K^*_0(1430) \to K \pi)} = 0.092 \pm 0.025^{+0.010}_{-0.025}$.
The \etaceta\ and $\Kstarz \to K \eta$ results correspond to the first 
  observations of these channels. The data also show evidence for \etactwopiz and first evidence for \etactwoeta.

\end{abstract}

\maketitle
\section{Introduction}
Charmonium decays, in particular $J/\psi$ radiative and hadronic decays, have been studied extensively~\cite{Kopke:1988cs,Bai:2003ww}.
One of the motivations for these studies is the search for non-$q \bar q$ mesons such as glueballs or molecular 
states that are predicted by
QCD to populate the low mass region of the hadron mass spectrum~\cite{glue}.
Recently, a search for exotic resonances was performed through Dalitz plot analyses of $\chi_{c1}$ states~\cite{cleo}.

Scalar mesons are still a puzzle in light-meson spectroscopy: there are too many states and they are not consistent with the quark model. 
In particular, the $f_0(1500)$ resonance, discovered in $\bar p p$ annihilations, has been interpreted as a scalar glueball~\cite{close}. 
However, no evidence for the $f_0(1500)$ state has been found in charmonium decays. 
Another glueball candidate is the $f_0(1710)$ discovered in radiative $J/\psi$ decays.
Recently, $f_0(1500)$ and $f_0(1710)$ signals have been incorporated
in a Dalitz plot analysis of $B \to 3K$ decays~\cite{babar_3k}. 
Charmless $B \to K X$ decays could show enhanced gluonium production~\cite{phen}. 
Another puzzling state is the $K^*_0(1430)$ resonance, never observed as a clear peak in the $K \pi$ mass spectrum. In the description of the scalar amplitude in 
$K \pi$ scattering, the $K^*_0(1430)$ resonance is added coherently to an effective-range
description of the low-mass  $K \pi$ system in such a way that the net amplitude
actually decreases rapidly at the resonance mass. 
The $K^*_0(1430)$  parameter values were measured by the LASS experiment in the reaction $\Km p \to \Km \pip n$~\cite{lass_kpi};
the corrected $\mathcal{S}$-wave amplitude representation is given explicitly in Ref.~\cite{babar_z}. 
In the present analysis, we study three-body \etac decays to pseudoscalar mesons and obtain results that are relevant to several issues
in light-meson spectroscopy.

Many \etac and \etactwo\ decay modes remain unobserved, while others have been studied with very limited statistical precision.
In particular, the branching fraction for the decay mode $\etac \to \Kp \Km \eta$ has been measured by the BESIII experiment based on a fitted 
yield of only $6.7\pm 3.2$ events~\cite{bes3}.
No Dalitz plot analysis has been performed on $\eta_c$ three-body decays. 

We describe herein a study of the \kketa\ and \kkpiz\ systems produced in two-photon interactions.
Two-photon events in which at least one of the interacting photons is not quasi-real are
strongly suppressed by the selection 
criteria described below. This implies that the allowed $J^{PC}$ values of
any produced resonances are $0^{\pm+}$, $2^{\pm+}$, $3^{++}$, $4^{\pm+}$...~\cite{Yang}.       
Angular momentum conservation, parity conservation, and charge conjugation
invariance imply that these quantum numbers also apply to
the final state except that the \kketa\ and \kkpiz\ states cannot be in a $J^P=0^+$ state.

This article is organized as follows. In Sec.\ II, a brief description of the
\babar\ detector is given. Section III is devoted to the event reconstruction and data selection. In Sec.\ IV, we describe the study of efficiency and resolution,
while in Sec.\ V the mass spectra are presented. Section VI is devoted to the measurement of the branching ratios, while Sec.\ VII describes the Dalitz plot analyses. In Sec.\ VIII, we report the measurement of the $K^*_0(1430)$ branching ratio, in Sec.\ IX we discuss its implications for the pseudoscalar meson mixing angle, and in Sec.\ X we summarize the results. 

\section{The \babar\ detector and dataset}

The results presented here are based on data collected
with the \babar\ detector 
at the PEP-II asymmetric-energy $e^+e^-$ collider
located at SLAC and correspond 
to an integrated luminosity of 519~\invfb~\cite{luminosity} recorded at
center-of-mass energies at and near the $\Upsilon (nS)$ ($n=2,3,4$)
resonances. 
The \babar\ detector is described in detail elsewhere~\cite{BABARNIM}.
Charged particles are detected, and their
momenta are measured, by means of a five-layer, double-sided microstrip detector,
and a 40-layer drift chamber, both operating  in the 1.5~T magnetic 
field of a superconducting
solenoid. 
Photons are measured and electrons are identified in a CsI(Tl) crystal
electromagnetic calorimeter. Charged-particle
identification is provided by the measurement of specific energy loss in
the tracking devices, and by an internally reflecting, ring-imaging
Cherenkov detector. Muons and \KL\ mesons are detected in the
instrumented flux return  of the magnet.
Monte Carlo (MC) simulated events~\cite{geant}, with sample sizes 
more than 10  times larger than the corresponding data samples, are
used to evaluate signal efficiency and to determine background features. 
Two-photon events are simulated  using the GamGam MC
generator~\cite{BabarZ}.

\section{Event Reconstruction and Data Selection}

In this analysis, we select events in which the $e^+$ and $e^-$  beam particles are scattered  
at small angles and are undetected in the final state. 
We study the following reactions
\begin{equation}
\gamma \gamma \to \Kp \Km \eta, (\eta \to \gamma \gamma),
\end{equation}
\begin{equation}
\gamma \gamma \to \Kp \Km \eta, (\eta \to \pip \pim \piz),
\end{equation}
and
\begin{equation}
\gamma \gamma \to \Kp \Km \piz.
\end{equation}
For reactions (1) and (3), we consider only events for which the number of well-measured charged-particle tracks with
transverse momenta greater than 0.1~\gevc\ is exactly equal to two.
For reaction (2), we require the number of well-measured charged-particle tracks to be exactly equal to four.
The charged-particle tracks are fit to a common vertex with the requirements that they
originate from the interaction region and that the $\chi^2$ probability of the vertex fit be greater than 0.1\%. 
We observe prominent \etac signals in all three reactions and improve the signal-to-background ratio using the data, in particular the $c \bar c$ \etac resonance. 
In the optimization procedure, we retain only selection criteria that do not remove significant \etac signal.
For the reconstruction of $\piz\to\gamma\gamma$ decays, we require the energy of the less-energetic photon to be
greater than 30 \mev\ for reaction~(2) and 50 \mev\ for reaction~(3).
For $\eta \to \gg$ decay, we require the energy of the less
energetic photon to be greater than 100~\mev.
Each pair of $\gamma$'s is kinematically fit to a \piz\ or $\eta$ hypothesis
requiring it to emanate from the primary vertex of the event, and with the diphoton
mass constrained to the nominal $\pi^0$ or $\eta$ mass, respectively~\cite{PDG}.
Due to the presence of soft-photon background, we do not impose a veto on the presence of additional photons in the final state.
For reaction (1), we require the presence of exactly one $\eta$ candidate in each event and discard events having additional \piz's decaying to $\gamma$'s with energy greater than 70 \mev. For reaction (3), we accept no more than two \piz \ candidates in the event.

In reaction (2), the $\eta$ is reconstructed by
combining two oppositely charged tracks identified as pions with each of the \piz \ candidates in the event. The
$\eta$ signal mass region is defined as $541<m(\pip\pim\piz)<554~\mevcc$. The momentum three-vectors of the the final-state pions 
are combined
and the energy of the $\eta$ candidate is computed using the nominal $\eta$ mass. 
According to tests with simulated events, this method
improves the \kketa mass resolution. We check for possible background from the reaction $\gamma \gamma \to \Kp \Km \pip \pim \piz$~\cite{kkpipipi0} using $\eta$ sideband regions and find it to be consistent with zero.
Background arises mainly from random combinations of particles from
\epem\ annihilation, from other two-photon processes, and from events with initial-state photon radiation (ISR). The ISR 
background is dominated by $J^{PC}=1^{--}$ resonance production~\cite{isr}.
We discriminate against \kketa\ (\kkpiz) events produced via ISR by requiring $\mm\equiv(p_{\epem}-p_{\mathrm{rec}})^2>10$~(GeV$^2$/$c^4$, where
$p_{\epem}$ is the four-momentum of the initial state and $p_{\mathrm{rec}}$ is the four-momentum of the \kketa\ (\kkpiz\ ) system. This requirement also removes a large fraction of a residual \jpsi contribution.

Particle identification is used in two different ways. For reaction (2), with four charged particles in the final state,  we require two oppositely charged particles to be loosely identified as kaons and 
the other two tracks to be consistent with pions. For reactions (1) and (3), with only two charged particles in the final state, we loosely identify one kaon and require that neither track be a well-identified pion, electron, or muon. 
We define \pt\ as the magnitude of the vector sum of the transverse momenta, in the \epem\ rest frame, of the final-state particles with respect to the beam axis.
Since well-reconstructed two-photon events are expected to have low
values of \pt, we require $\pt<0.05~\gevc$.
Reaction (3) is affected by background from the reaction $\gamma \gamma \to \Kp \Km$ where soft photon background simulates the presence
of a low momentum \piz. We reconstruct this mode and reject events having a $\gamma \gamma \to \Kp \Km$ candidate with $\pt < 0.1~\gevc$.

\begin{figure}[hb]
\begin{center}
\includegraphics[width=7cm]{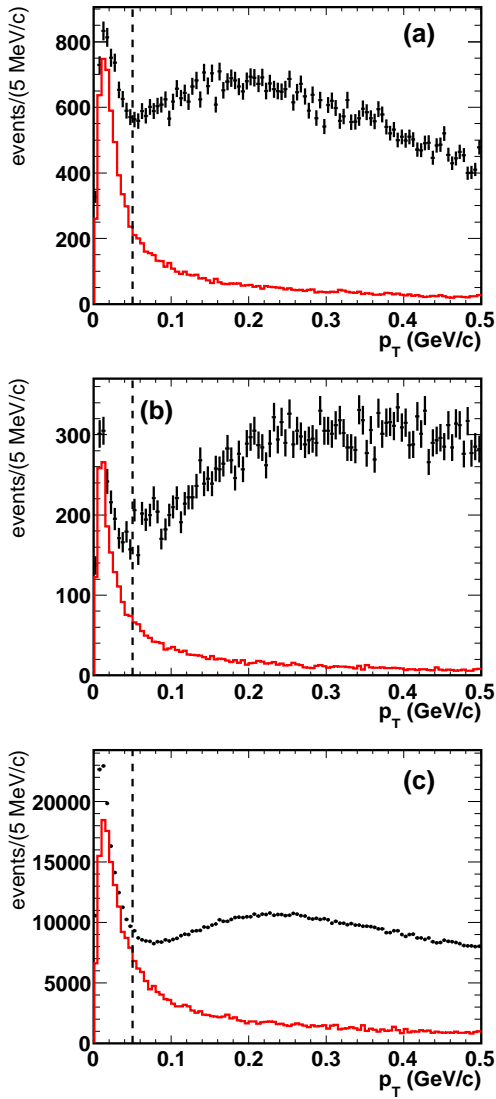}
\caption{Distributions of \pt\ for (a) $\gamma \gamma \to \Kp \Km \eta \ (\eta \to \gamma \gamma)$, (b) $\gamma \gamma \to \Kp \Km \eta \ (\eta \to \pip \pim \piz)$, and (c) $\gamma \gamma \to \Kp \Km \piz$. In each figure the data are shown as points with error bars,
and the MC simulation is shown as a histogram; the vertical dashed line indicates the selection applied to isolate two-photon events.}
\label{fig:fig1}
\end{center}
\end{figure}

Figure~\ref{fig:fig1} shows the  measured \pt\ distribution for each of the three reactions in comparison to the corresponding  \pt\ distribution obtained from simulation.
A peak at low \pt\ is observed in all three distributions indicating
the presence of the two-photon process. The shape of the peak agrees well with that seen in the MC simulation.

\section{Efficiency and resolution}

To compute the efficiency, \etac and \etactwo MC signal events for the different channels are generated using a detailed detector simulation~\cite{geant} in which the \etac and \etactwo mesons decay uniformly in phase space.
These simulated events are reconstructed and analyzed in the same manner as data. The efficiency is computed as the ratio of reconstructed to
generated events. Due to the presence of long tails in the Breit-Wigner (BW) representation of the resonances, we apply 
selection criteria to restrict the generated events to the \etac and \etactwo mass regions. 
We express the efficiency as a function of the $m(\Kp \Km)$ mass and $\cos \theta$, where $\theta$ is the angle in the $K^+K^-$ 
rest frame between the directions of the \Kp and the boost from the $K^+K^-\eta$ or $K^+K^-\pi^0$ rest frame.
To smooth statistical fluctuations, this efficiency is then parameterized as follows.

\begin{figure}[h]
\begin{center}
\includegraphics[width=7cm]{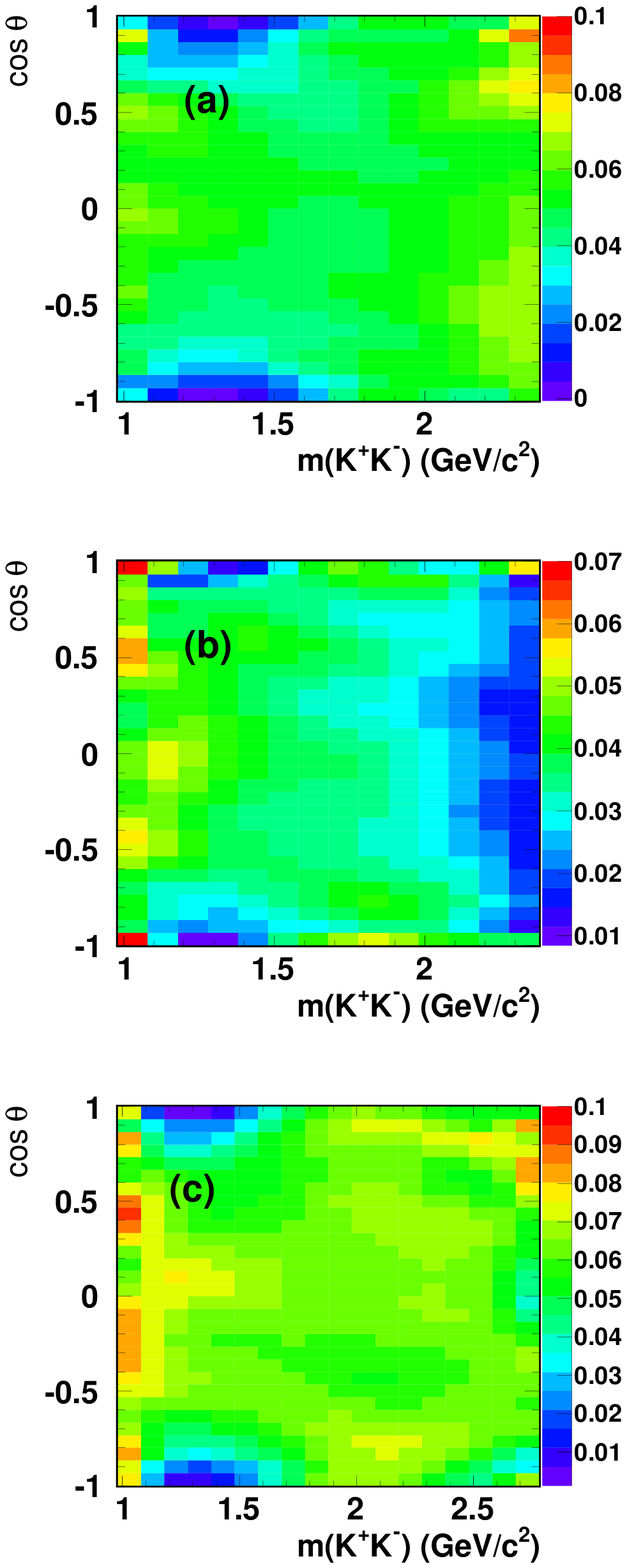}
\caption{Fitted detection efficiency in the $\cos \theta \  vs. \ m(\Kp \Km)$ plane for (a) \etaceta (\etagg), 
(b) \etaceta (\etapipipi), and (c) \etacpiz. Each bin shows the average value of the fit in that region.}
\label{fig:fig2}
\end{center}
\end{figure}

First we fit the efficiency as a function of  $\cos \theta$ in separate intervals of $m(\Kp \Km)$, in terms 
of Legendre polynomials up to $L=12$:
\begin{eqnarray}
\epsilon(\cos\theta) = \sum_{L=0}^{12} a_L(m) Y^0_L(\cos\theta),
\end{eqnarray}
where $m$ denotes $\Kp \Km$ invariant mass.
For each value of $L$, we fit the mass dependent coefficients $a_L(m)$ with a seventh-order polynomial in $m$.
Figure~\ref{fig:fig2} shows the resulting fitted efficiency $\epsilon(m,\cos \theta)$ for each of the three reactions. We observe a significant decrease in
efficiency for $\cos\theta \sim \pm 1$ and $1.1<m(\Kp \Km)<1.5~\gevcc$ due to the impossibility of reconstructing low-momentum kaons (p$<$200 \mevc in the laboratory frame) which have experienced significant energy loss in the beampipe and inner-detector material.
The efficiency decrease at high $m$
for \etaceta (\etapipipi) (Fig.~\ref{fig:fig2}(b)) results from the loss of a low-momentum \piz\ from the $\eta$ decay.

\begin{figure}[h]
\begin{center}
\includegraphics[width=6cm]{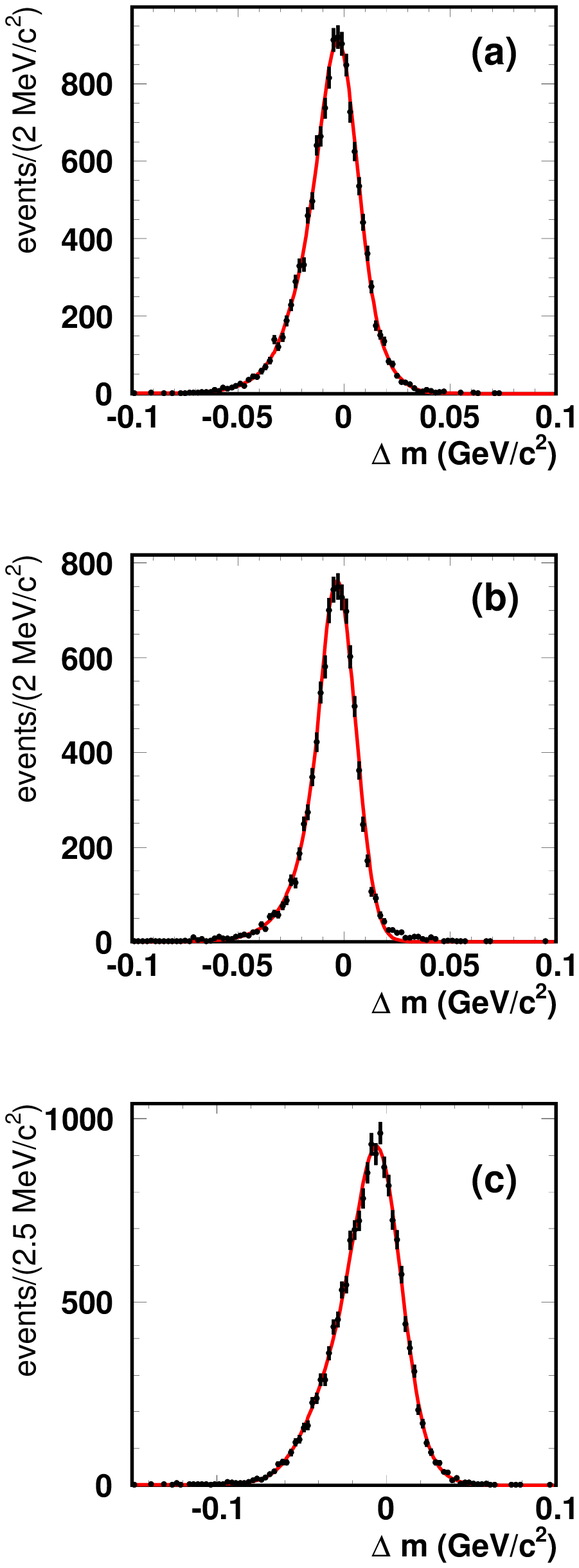}
\caption{MC mass resolution for (a) \etaceta (\etagg), 
(b) \etaceta (\etapipipi), and (c) \etacpiz. The curves represent the fits described in the text.}
\label{fig:fig3}
\end{center}
\end{figure}

The mass resolution, $\Delta m$, is measured as the difference between the generated and reconstructed \kketa\ or \kkpiz\ invariant-mass values.
Figure~\ref{fig:fig3} shows the $\Delta m$ distribution for each of the \etac signal regions; these deviate from Gaussian line
shapes due to a low-energy tail caused by the response of the CsI calorimeter to photons. We fit the distribution for the \kketa\ ($\eta \to \pipipi$) final state to a Crystal Ball function~\cite{cb}, and those for the \kketa ($\eta \to \gg$) and \kkpiz\ final states to a
sum of a Crystal Ball function and a Gaussian function. 
The root-mean-squared values are 15, 14, and 21 \mevcc \ at the $\eta_c$ mass, and 18, 15, and 24 \mevcc \ at the \etactwo\ mass, for the \kketa ($\eta \to \gg$), \kketa ($\eta \to \pipipi$), and \kkpiz final states, respectively. 

\section{Mass spectra}

\begin{figure*}[ht]
\begin{center}
\includegraphics[width=18cm]{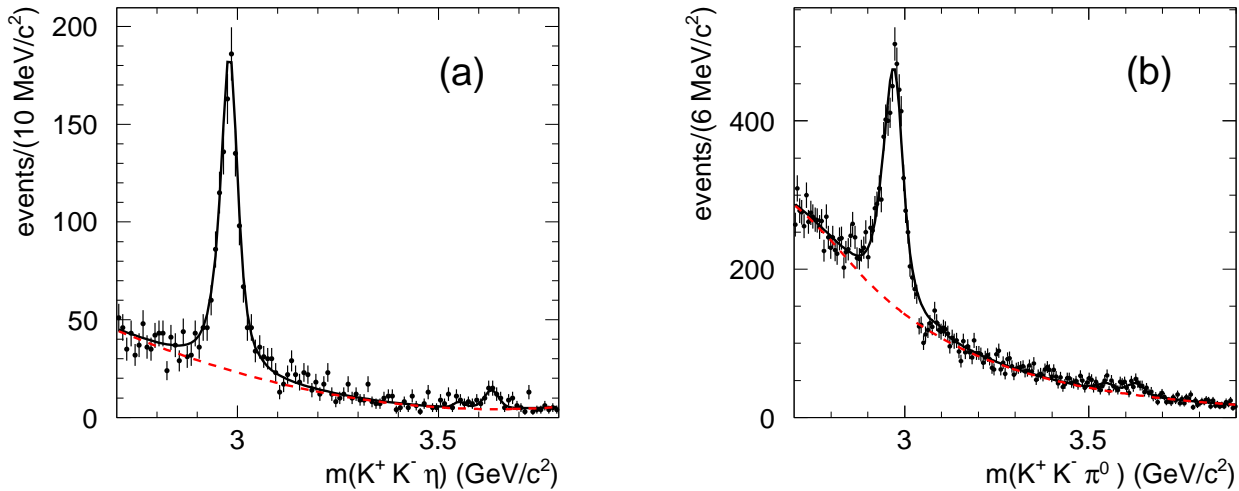}
\caption{(a) The \kketa\ mass spectrum summed over the two $\eta$ decay modes. (b) The \kkpiz  mass spectrum. In each figure, the solid curve shows the total fitted function
and the dashed curve shows the fitted background contribution.}
\label{fig:fig4}
\end{center}
\end{figure*}

Figure~\ref{fig:fig4}(a) shows the \kketa\ mass spectrum, summed over the two $\eta$ decay modes, before applying the efficiency correction. There are 2950 events in the mass region between
2.7 and 3.8 \gevcc, of which 73\% are from the \etagg\ decay mode and 27\% are from the \etapipipi decay mode. 
We observe a strong \etac signal and a small enhancement at the position of the \etactwo. The \etac signal-to-background ratio for each of the $\eta$ decay modes is approximately the same.
We perform a simultaneous fit to the \kketa mass spectra for the two $\eta$ decay modes. For each resonance, 
the mass and width are constrained to take the same fitted values in both distributions. Backgrounds are described by second-order polynomials, and each resonance is represented by a simple Breit-Wigner function convolved with the corresponding resolution function. In addition, we include a signal function for the $\chi_{c2}$ resonance with parameters fixed to their PDG values~\cite{PDG}. Figure~\ref{fig:fig4}(a) shows the fit result, and Table~\ref{tab:mass} summarizes the \etac and \etactwo parameter values. We have only a weak constraint
on the \etactwo\ width and so fix its value to 11.3 \mev~\cite{PDG}.

The \kkpiz mass spectrum is shown in Fig.~\ref{fig:fig4}(b). There are 23\,720 events in the mass region between 2.7 and 3.9 \gevcc. We observe a strong \etac signal and a small signal at the position of the \etactwo on top of a sizeable background. We perform a fit to the \kkpiz mass spectrum
using the background function $B(m) = e^{a_1m+a_2m^2}$ for $m<m_0$ and $B(m) = e^{b_0+b_1m+b_2m^2}$ for $m>m_0$, where $m=m(\Kp \Km \piz)$ and $a_i$, $b_i$, and $m_0$ are free parameters~\cite{babar_dj}.
The two functions and their first derivatives are required to be continuous at $m_0$, so that the resulting function has only four independent parameters.
In addition, we allow for the presence of a residual \jpsi contribution modeled as a simple Gaussian function. Its parameter values are fixed to those from a fit to the \kkpiz mass spectrum for the ISR data sample obtained requiring
$\lvert \mm \rvert<1 \ (\gevcc)^2$. Figure~\ref{fig:fig4}(b) shows the fit to the $\Kp \Km \piz$ mass spectrum, and Table~\ref{tab:mass} summarizes the resulting \etac and \etactwo parameter values.  

\begin{table}[ht]
\caption{Fitted \etac and \etactwo parameter values. The first uncertainty is statistical and the second is systematic.}
\label{tab:mass}
\begin{center}
\vskip -0.2cm
\begin{tabular}{lcc}
\hline
 \noalign{\vskip2pt}
Resonance & Mass (\mevcc)& $\Gamma$ (\mev) \cr
\hline
\hline
\etaceta & $2984.1 \pm 1.1 \pm 2.1$ & $34.8 \pm 3.1 \pm 4.0$  \cr
\etacpiz & $2979.8 \pm 0.8 \pm 3.5$ & $25.2 \pm 2.6 \pm 2.4$  \cr
\hline
\etactwoeta & $3635.1 \pm 5.8 \pm 2.1$ & 11.3 (fixed)  \cr
\etactwopiz & $3637.0 \pm 5.7 \pm 3.4$ & 11.3 (fixed) \cr
\hline
\end{tabular}
\end{center}
\end{table}

The following systematic uncertainties are considered. The background uncertainty contribution is estimated by replacing each function by a third-order polynomial. The mass scale uncertainty is estimated from fits to the \jpsi signal in ISR events. In the case of $\eta_c \to  \Kp \Km \eta$, we perform independent
fits to the mass spectra obtained for the two $\eta$ decay modes, and consider the mass difference as a measurement of systematic uncertainty. 
The different contributions are added in quadrature to obtain the values quoted in Table~\ref{tab:mass}.

\section{Branching ratios}

We compute the ratios of the branching fractions for \etac and \etactwo decays to the \kketa final state compared to the respective branching fractions to the \kkpiz final state. The ratios are computed as
\begin{equation}
\begin{split}
\calR = &\frac{\BR(\etac/\etactwo \to K^+ K^- \eta)}{\BR(\etac/\etactwo \to K^+ K^- \pi^0)} \\
  = &\frac{N_{K^+ K^- \eta}}{N_{K^+ K^- \pi^0}}\frac{\epsilon_{K^+ K^- \pi^0}}{\epsilon_{K^+ K^- \eta}}\frac{1}{\BR_{\eta}}.
\end{split}
\end{equation}
For each $\eta$ decay mode, $N_{K^+ K^- \eta}$ and $N_{K^+ K^- \pi^0}$ represent the fitted yields for \etac and \etactwo in the \kketa and \kkpiz mass spectra, $\epsilon_{K^+ K^- \eta}$ and $\epsilon_{K^+ K^- \pi^0}$ are the corresponding efficiencies, and $\BR_{\eta}$ indicates the particular $\eta$ branching fraction. The PDG values of the branching fractions are $(39.41\pm0.20)$\% and $(22.92 \pm 0.28)$\% for the $\eta \to \gamma \gamma$ and $\eta \to \pipipi$, respectively~\cite{PDG}.
We estimate $\epsilon_{K^+ K^- \eta}$ and $\epsilon_{K^+ K^- \pi^0}$ for the \etac signals by making use of the \mbox{2-D} efficiency 
functions described in Sec. IV and weighting each event by $1/\epsilon(m,\cos \theta)$. 
Due to the presence of non-negligible backgrounds in the \etac signals, which have different distributions in the
Dalitz plot, we perform a sideband subtraction by assigning a weight +1 to events in the signal region and
a negative weight to events in the sideband regions. The weight in the sideband regions is scaled down
to match the fitted \etac signal/background ratio.
To remove the dependence of the fit quality on the efficiency functions we make use of the unfitted efficiency distributions.
Due to the presence of a sizeable background for the \etactwo, we use the average efficiency value from the simulation.

We determine $N_{K^+K^-\eta}$ and $N_{K^+K^-\pi^0}$ for the \etac by performing fits to the \kketa and \kkpiz mass spectra. The width is extracted from the simultaneous fit to the \kketa mass spectra, and is fixed to this value in the fit to the \kkpiz mass spectrum. 
This procedure is adopted because the signal-to-background ratio at the peak is much better for the \kketa mode ($\sim$8:1 compared to $\sim$2:1 for the \kkpiz mode) while the residual $J/\psi$ contamination is much smaller.
The \etac and \etactwo mass values are determined from the fits.
For the \etactwo, we fix the width to 11.3 \mev~\cite{PDG}.  The resulting yields, efficiencies, measured branching ratios, and significances are reported in Table~\ref{tab:tab2}. The significances are evaluated as $N_s/\sigma_T$ where $N_s$ is the signal event yield and $\sigma_T$ is the 
total uncertainty obtained by adding the statistical and systematic contributions in quadrature.
\begin{table*}[ht]
\caption{Summary of the results from the fits to the \kketa and \kkpiz mass spectra. The table lists event yields, efficiency correction weights, resulting branching ratios and significances. For event yields, the first uncertainty is statistical and the second is systematic. In the evaluation of significances, systematic uncertainties are included. }
\label{tab:tab2}
\begin{center}
\vskip -0.2cm
\begin{tabular}{lcccc}
\hline
 \noalign{\vskip2pt}
Channel & Event yield & Weights &  $\calR$  & Significance \cr
\hline
\hline
 \noalign{\vskip2pt}
$\eta_c \to  K^+ K^-\piz$ & 4518 $\pm$ 131 $\pm$ 50 & 17.0 $\pm$ 0.7 & & 32~$\sigma$\cr
\hline
 \noalign{\vskip1pt}
$\eta_c \to  K^+ K^-\eta$ ($\eta \to \gamma \gamma$)& 853 $\pm$ 38 $\pm$ 11 & 21.3  $\pm$ 0.6 & & 21~$\sigma$ \cr
\hline
 \noalign{\vskip1pt}
$\BR(\eta_c \to  K^+ K^-\eta)/\BR(\eta_c \to  K^+ K^-\piz)$ & & & $ 0.602 \pm 0.032 \pm 0.065$ & \cr
\hline
 \noalign{\vskip1pt}
$\eta_c \to  K^+ K^-\eta$ ($ \eta \to \pip \pim \piz$) & 292 $\pm$ 20 $\pm$ 7 & 31.2  $\pm$ 2.1 & & 14~$\sigma$ \cr
\hline
 \noalign{\vskip1pt}
$\BR(\eta_c \to  K^+ K^-\eta)/\BR(\eta_c \to  K^+ K^-\piz)$ & & & $0.523 \pm 0.040 \pm 0.083$ & \cr
\hline
 \noalign{\vskip1pt}
$\eta_c(2S) \to  K^+ K^-\piz$ & 178 $\pm$ 29 $\pm$ 39 & 14.3 $\pm$ 1.3 & &  3.7~$\sigma$ \cr
$\eta_c(2S) \to K^+ K^-\eta $ & 47 $\pm$ 9 $\pm$ 3 & 17.4 $\pm$ 0.4 & &   4.9~$\sigma$ \cr
\hline
 \noalign{\vskip1pt}
$\BR(\eta_c(2S) \to  K^+ K^-\eta)/\BR(\eta_c(2S) \to  K^+ K^-\piz)$ & & & $0.82 \pm 0.21 \pm 0.27$ & \cr
\hline
 \noalign{\vskip1pt}
\chictwotopiz &  88 $\pm$ 27 $\pm$ 23 & & & 2.5~$\sigma$ \cr
\hline
 \noalign{\vskip1pt}
\chictwotoeta &  2 $\pm$ 5 $\pm$ 2 & & & 0.0~$\sigma$ \cr
\hline
\end{tabular}
\end{center}
\end{table*}

We calculate the weighted mean of the \etac branching-ratio estimates for the two $\eta$ decay modes and obtain
\begin{equation}
\calR(\etac) = \frac{\BR(\etac \to  K^+ K^-\eta)}{\BR(\etac \to  K^+ K^-\piz)} = 0.571 \pm 0.025 \pm 0.051,
\end{equation}
which is consistent with the BESIII measurement of $0.46 \pm  0.23$~\cite{bes3}.
Since the sample size for $\etactwo \to  K^+ K^-\eta$ decays with \etapipipi is small, we use only the \etagg\ decay mode, and 
obtain
\begin{equation}
\calR(\etactwo) = \frac{\BR(\etactwo \to  K^+ K^-\eta)}{\BR(\etactwo \to  K^+ K^-\piz)} = 0.82 \pm 0.21 \pm 0.27.
\end{equation}
In evaluating $\calR(\etac)$ for the $\eta \to \gg$ decay mode, we note that the number of charged-particle tracks
and $\gamma$'s is the same in the numerator and in the denominator of the ratio, so that several systematic uncertainties cancel.
Concerning the contribution of the $\eta \to \pipipi$ decay, we find systematic uncertainties related to the difference in the number of
charged-particle tracks to be negligible.
We consider the following sources of systematic uncertainty.
We modify the \etac width by fixing its value to the PDG value~\cite{PDG}. 
We modify the background model by using fourth-order polynomials or exponential functions.
The uncertainty due to the efficiency weight
is evaluated by computing 1000 new weights obtained by randomly modifying the weight in each cell of the 
$\epsilon(m(\Kp \Km),\cos \theta)$ plane according to its
statistical uncertainty. 
The widths of the resulting Gaussian distributions yield the estimate of the systematic uncertainty for the efficiency weighting procedure. These values are reported as the weight uncertainties in Table~\ref{tab:tab2}.

\section{Dalitz plot analyses}

We perform Dalitz plot analyses of the \kketa and \kkpiz systems in the \etac mass region using  unbinned maximum likelihood fits.
The likelihood function is written as
\begin{eqnarray}
\mathcal{L} = \nonumber\\
 \prod_{n=1}^N&\bigg[&f_{\rm sig}(m_n) \cdot \epsilon(x'_n,y'_n)\frac{\sum_{i,j} c_i c_j^* A_i(x_n,y_n) A_j^*(x_n,y_n)}{\sum_{i,j} c_i c_j^* I_{A_i A_j^*}} \nonumber\\
& &+(1-f_{\rm sig}(m_n))\frac{\sum_{i} k_iB_i(x_n,y_n)}{\sum_{i} k_iI_{B_i}}\bigg]
\end{eqnarray}
\noindent where
\begin{itemize}
\item $N$ is the number of events in the signal region;
\item for the $n$-th event, $m_n$ is the \kketa or the \kkpiz invariant mass;
\item for the $n$-th event, $x_n=m^2(K^+ \eta)$, $y_n=m^2(K^- \eta)$ for $\kketa$; $x_n=m^2(K^+ \piz)$, $y_n=m^2(K^- \piz)$ for $\kkpiz$; 
\item $f_{\rm sig}$ is the mass-dependent fraction of signal obtained from the fit to the \kketa or \kkpiz mass spectrum;
\item for the $n$-th event, $\epsilon(x'_n,y'_n)$ is the efficiency parameterized as a function $x'_n=m(\Kp \Km)$ and $y'_n=\cos \theta$ (see Sec. IV);
\item for the $n$-th event, the $A_i(x_n,y_n)$ describe the complex signal-amplitude contributions;
\item $c_i$ is the complex amplitude of the $i-$th signal component; the $c_i$ parameters are allowed to vary during the fit process;
\item for the $n$-th event, the $B_i(x_n,y_n)$ describe the background probability-density functions assuming that interference between signal and background amplitudes can be ignored;
\item $k_i$ is the magnitude of the $i-$th background component; the $k_i$ parameters are obtained by fitting the sideband regions;
\item $I_{A_i A_j^*}=\int A_i (x,y)A_j^*(x,y) \epsilon(m(\Kp \Km),\cos \theta)\ {\rm d}x{\rm d}y$ and 
$I_{B_i}~=~\int B_i(x,y) {\rm d}x{\rm d}y$ are normalization
 integrals; numerical integration is performed on phase space generated events.
\end{itemize}
Amplitudes are parameterized as described in Refs.~\cite{asner} and~\cite{ds}.
The efficiency-corrected fractional contribution $f_i$ due to resonant or nonresonant contribution $i$ is defined as follows:
\begin{equation}
f_i = \frac {|c_i|^2 \int |A_i(x_n,y_n)|^2 {\rm d}x {\rm d}y}
{\int |\sum_j c_j A_j(x,y)|^2 {\rm d}x {\rm d}y}.
\end{equation}
The $f_i$ do not necessarily sum to 100\% because of interference effects. The uncertainty for each $f_i$ is evaluated by propagating the full covariance matrix obtained from the fit. 

\subsection{Dalitz plot analysis of {\boldmath$\protect \etaceta$}\ }

\begin{figure}[h]
\begin{center}
\includegraphics[width=9cm]{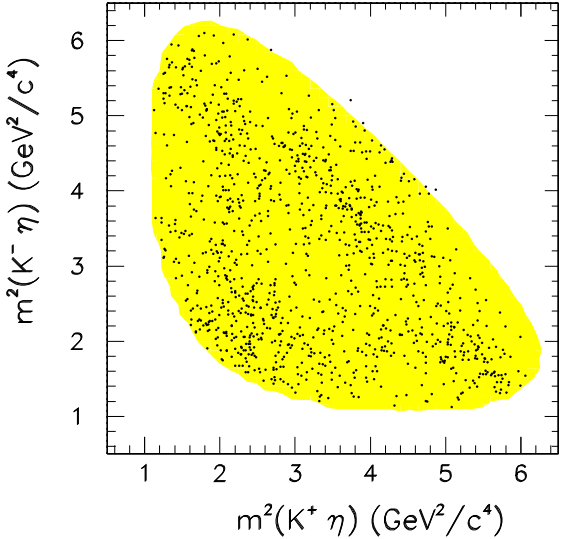}
\caption{Dalitz plot for the \etaceta\ events in the signal region. The shaded area denotes the accessible kinematic region.}
\label{fig:fig5}
\end{center}
\end{figure}

\begin{figure}[h]
\begin{center}
\includegraphics[width=7cm]{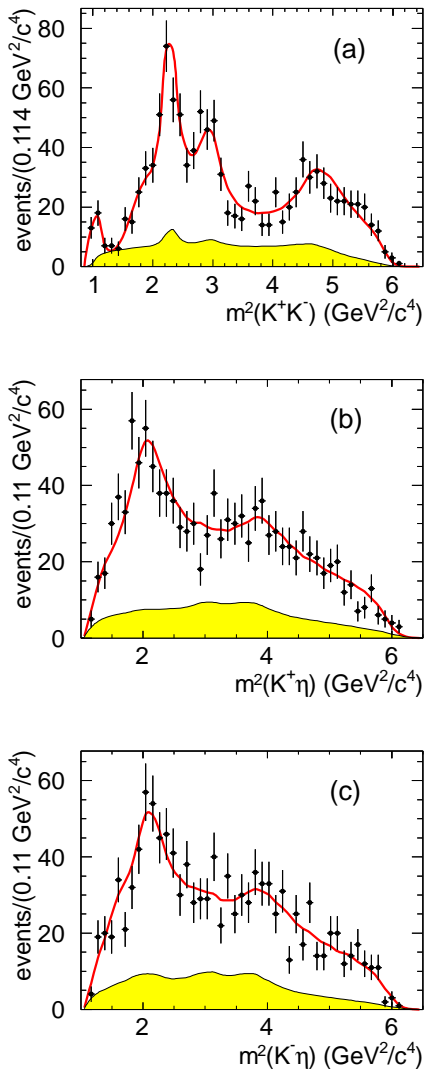}
\caption{The \etaceta\ Dalitz plot projections. The superimposed curves result from the Dalitz plot analysis described in the text. The shaded regions show the
background estimates obtained by interpolating the results of the Dalitz plot analyses of the sideband regions.
}
\label{fig:fig6}
\end{center}
\end{figure}

We define the \etac signal region as the range 2.922-3.036~\gevcc. This region contains 1161 events with (76.1 $\pm$ 1.3)\% purity, defined as 
$S/(S+B)$ where $S$ and $B$ indicate the number of signal and background events, respectively, as determined from the fit (Fig.~\ref{fig:fig4}(a)).
Sideband regions are defined as the ranges 2.730-2.844~\gevcc \ and 3.114-3.228~\gevcc, respectively.
Figure~\ref{fig:fig5} shows the Dalitz plot for the \etac signal region and Fig.~\ref{fig:fig6} shows the Dalitz plot projections.

We observe signals in the $\Kp \Km$ projections corresponding to the $f_0(980)$, $f_0(1500)$, $f_0(1710)$, and $f_0(2200)$ states. We also observe a broad signal in the 1.43 \gevcc \ mass
region in the $\Kp \eta$ and $\Km \eta$ projections.

In describing the Dalitz plot, we note that amplitude contributions to the $\Kp \Km$ system must have isospin zero 
in order to satisfy overall 
isospin conservation in \etac decay. In addition, amplitudes of the form $K^*\bar K$ must be symmetrized as $(K^{*+}\Km + K^{*-}\Kp)/\sqrt{2}$
so that the decay conserves C-parity. 
For convenience, these amplitudes are denoted by $K^{*+}K^-$ in the following.

\begin{table}[h]
\caption{Results of the Dalitz plot analysis of the $\eta_c \to  \Kp \Km \eta$ channel.}
\label{tab:signal_etakk}
\begin{center}
\vskip -0.2cm
\begin{tabular}{lr@{}c@{}lr@{}c@{}l}
\hline
 \noalign{\vskip2pt}
Final state & \multicolumn{3}{c}{Fraction \%} & \multicolumn{3}{c}{Phase (radians)}\cr
\hline
\hline
 \noalign{\vskip2pt}
$f_0(1500) \eta$  & 23.7 $\pm$ & \, 7.0 $\pm$ & \, 1.8 & 0.\cr
$f_0(1710) \eta$  & 8.9 $\pm$ & \, 3.2 $\pm$ & \, 0.4 &  2.2 $\pm$ & \, 0.3 $\pm$ & \, 0.1 \cr
$K^{*}_0(1430)^+ K^-$ & 16.4 $\pm$ & \, 4.2 $\pm$ & \, 1.0 &  2.3 $\pm$ & \, 0.2 $\pm$ & \, 0.1\cr
$f_0(2200) \eta$  &11.2 $\pm$ & \, 2.8 $\pm$ & \, 0.5 & 2.1  $\pm$ & \, 0.3 $\pm$ & \, 0.1 \cr
$K^{*}_0(1950)^+ K^-$ & 2.1 $\pm$ & \, 1.3 $\pm$ & \, 0.2 & -0.2 $\pm$ & \, 0.4 $\pm$ & \, 0.1\cr
$f_2'(1525) \eta$ & 7.3 $\pm$ & \, 3.8 $\pm$ & \, 0.4 & 1.0 $\pm$ & \, 0.1 $\pm$ & \, 0.1 \cr
$f_0(1350) \eta$  & 5.0 $\pm$ & \, 3.7 $\pm$ & \, 0.5 & 0.9 $\pm$ & \, 0.2 $\pm$ & \, 0.1 \cr
$f_0(980) \eta$  & 10.4 $\pm$ & \, 3.0 $\pm$ & \, 0.5 & -0.3 $\pm$ & \, 0.3 $\pm$ & \, 0.1 \cr
{\it NR} & 15.5 $\pm$ & \, 6.9 $\pm$ & \, 1.0 & -1.2 $\pm$ & \, 0.4 $\pm$ & \, 0.1\cr
 \noalign{\vskip1pt}
\hline
 \noalign{\vskip1pt}
Sum & 100.0 $\pm$ & \,11.2  $\pm$ & \, 2.5 & \cr
$\chi^2/\nu$ & & \, 87/65 & \,& \cr
\hline
\end{tabular}
\end{center}
\end{table}

The $f_0(980)$ is parameterized as in a \babar\ Dalitz plot analysis of $D_s^+ \to \Kp \Km \pip$ decay~\cite{ds}. 
For the $f_0(1430)$ we use the BES parameterization~\cite{bes}. 
For the  $K^{*}_0(1430)$, we use our results from the Dalitz plot analysis (see Sec. VII.C), since the individual
measurements of the mass and width considered for the PDG average values~\cite{PDG} show
a large spread for each parameter. 
The non-resonant ({\it NR}) contribution is parameterized as an amplitude that is constant in magnitude and phase
over the Dalitz plot. The $f_0(1500) \eta$ amplitude is taken as the reference amplitude, and so its phase is set to zero.
The test of the fit quality is performed by computing a two-dimensional \mbox{(2-D)} $\chi^2$ over the Dalitz plot. 

We first perform separate fits to the \etac sidebands using a list of incoherent sum of amplitudes. We find significant contributions from the $f_2'(1525)$, $f_0(2200)$, $K^*_3(1780)$, and $K^*_0(1950)$ resonances, as well as from an incoherent uniform background.
The resulting amplitude fractions are interpolated 
into the \etac signal region and normalized to yield the fitted purity. Figure~\ref{fig:fig6} shows the projections of the estimated background contributions as shaded distributions.

For the description of the \etac signal, amplitudes are added one by one to ascertain the associated increase of the likelihood value and decrease of the \mbox{2-D} $\chi^2$. 
Table~\ref{tab:signal_etakk} summarizes the fit results for the amplitude fractions and phases. We note that the $f_0(1500) \eta$ amplitude provides the largest contribution. We also observe important contributions from the $K^{*}_0(1430)^+ K^-$, $f_0(980) \eta$, $f_0(2200) \eta$, and $f_0(1710) \eta$ channels. In addition, the fit requires a sizeable {\it NR} contribution. The sum of the fractions
for this \etac decay mode is consistent with 100\%.

We test the statistical significance of the $K^{*}_0(1430)^+ \Km$ contribution by removing it from the list of amplitudes.
We obtain a change of the negative log likelihood $\Delta(-2{\ln} \mathcal{L})$=+107 and an increase of the $\chi^2$ on the Dalitz plot $\Delta \chi^2$=+76 for the reduction by 2 parameters. This corresponds to a statistical significance of 10.3 standard deviations. We obtain the first observation of the $K^{*}_0(1430)^{\pm} \to K^{\pm} \eta$ decay mode.

We test the quality of the fit by examining a large sample of MC events at the generator level weighted 
by the likelihood fitting function and by the efficiency. These events are used to
compare the fit result to the Dalitz plot and its projections with proper normalization. The latter comparison is shown in Fig.~\ref{fig:fig6}, and good agreement is obtained for
all projections. We make use of these weighted events to compute a \mbox{2-D} $\chi^2$ over the Dalitz plot. For this purpose, we divide the Dalitz plot into a number of cells such that the expected population in each cell is at least eight events. We compute 
$\chi^2 = \sum_{i=1}^{N_{cells}} (N^i_{obs}-N^i_{exp})^2/N^i_{exp}$, where $N^i_{obs}$ and $N^i_{exp}$ are event yields from data and simulation, respectively. 
Denoting by $n\ (=16)$ the number of free parameters in the fit, we obtain $\chi^2/\nu=87/65$ ($\nu=N_{cells}-n$), 
which indicates that the description of the data is adequate.

\begin{figure*}[h]
\begin{center}
\includegraphics[width=16cm]{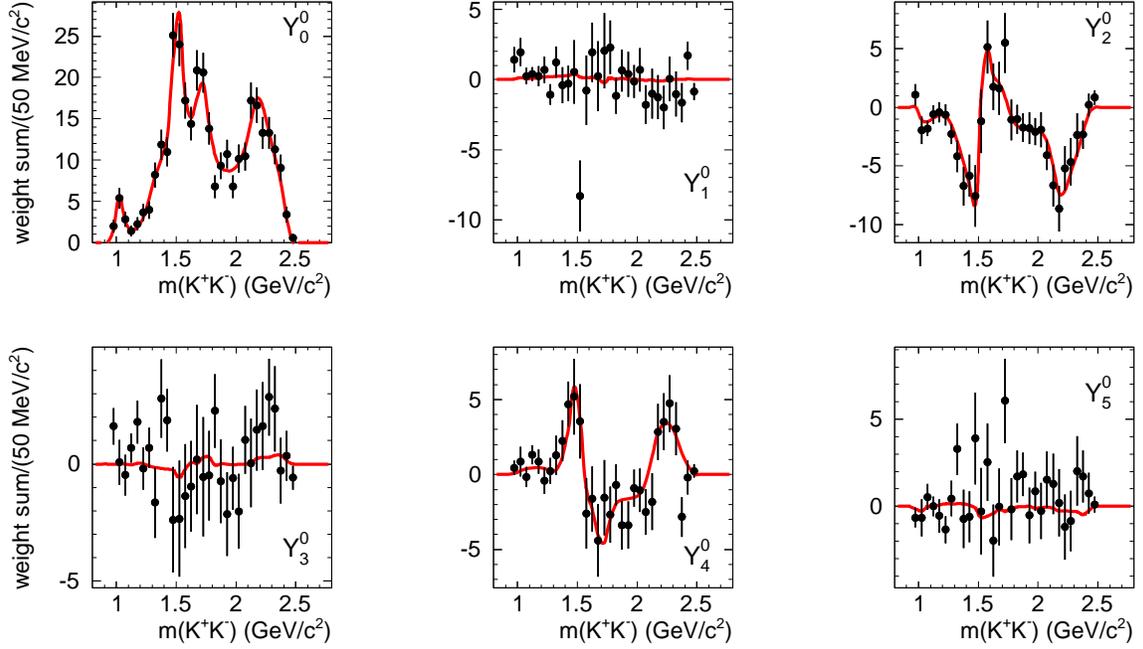}
\caption{Legendre polynomial moments for \etaceta as a function of $\Kp \Km$ mass. The superimposed curves result from the Dalitz plot analysis described in the text.}
\label{fig:fig7}
\end{center}
\end{figure*}

\begin{figure*}[h]
\begin{center}
\includegraphics[width=16cm]{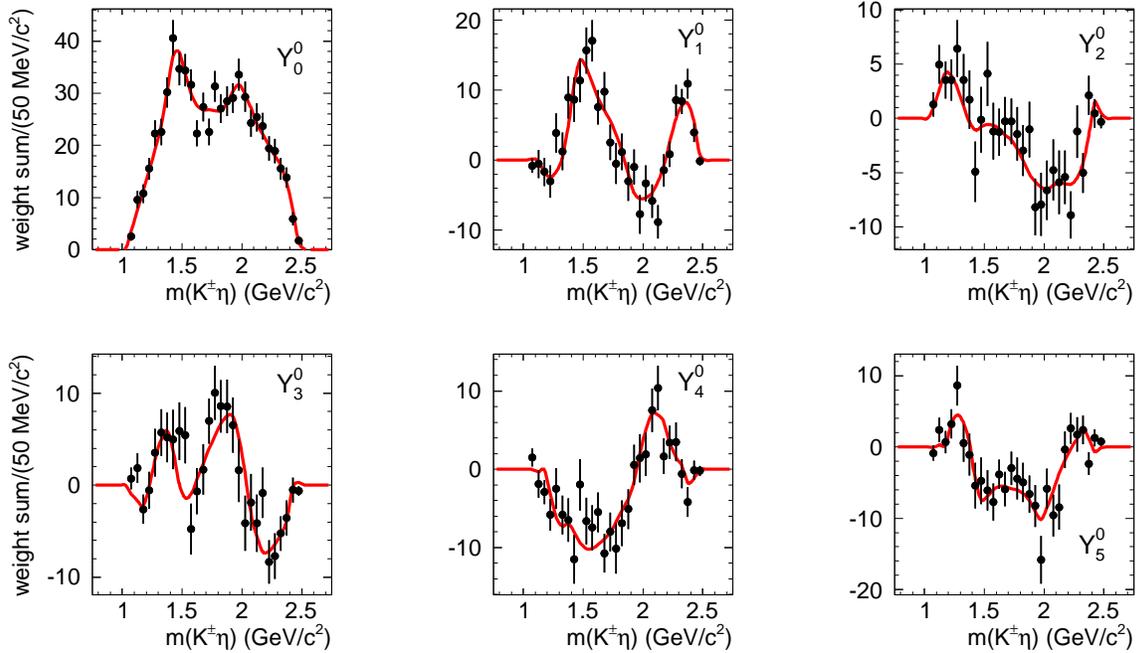}
\caption{Legendre polynomial moments for \etaceta as a function of $\Kpm \eta$ mass. The superimposed curves result from the Dalitz plot analysis described in the text. The corresponding $\Kp \eta$ and $\Km \eta$ distributions are combined.}
\label{fig:fig8}
\end{center}
\end{figure*}

We compute the uncorrected Legendre polynomial moments $\langle Y^0_L \rangle$ in each  $\Kp \Km$ and $\eta \Kpm$  mass interval by weighting each event by the relevant $Y^0_L(\cos \theta)$ function. These distributions are shown in Figs.~\ref{fig:fig7} and~\ref{fig:fig8}. We also compute the expected Legendre polynomial moments from the weighted MC events and compare with the experimental distributions. We observe good agreement for all the distributions, which indicates that the fit is able to reproduce the local structures apparent in the Dalitz plot.

Systematic uncertainty estimates for the fractions and relative phases are computed in two different ways: 1) the purity function is scaled up and down by its statistical uncertainty, and 2) the parameters of each resonance contributing to the decay are modified within one standard deviation  
of their uncertainties in the PDG average. The two contributions are added in quadrature.

\subsection{Dalitz plot analysis of {\boldmath$\protect \etacpiz$}\ }

\begin{figure}[h]
\begin{center}
\includegraphics[width=8cm]{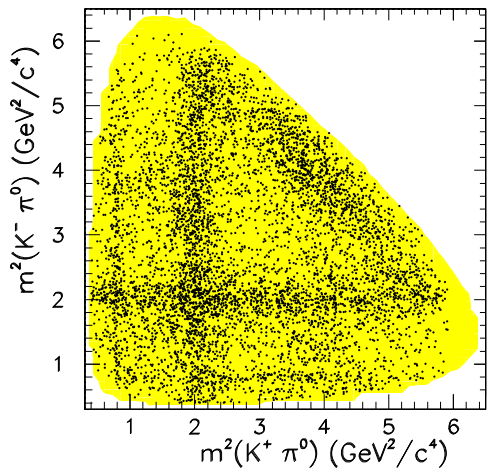}
\caption{Dalitz plot for the events in the \etacpiz\ signal region. The shaded area denotes the accessible kinematic region.}
\label{fig:fig9}
\end{center}
\end{figure}

\begin{figure}[h]
\begin{center}
\includegraphics[width=7cm]{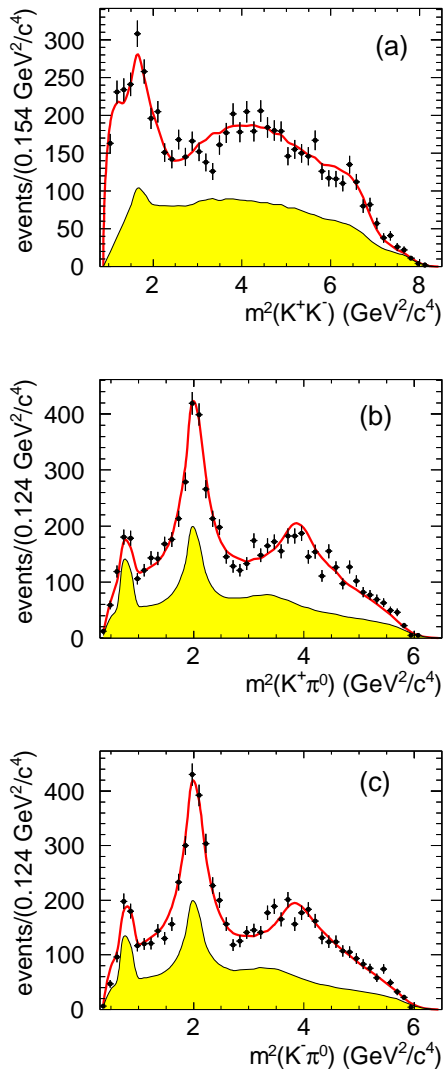}
\caption{The \etacpiz\ Dalitz plot projections. The superimposed curves result from the Dalitz plot analysis described in the text. The shaded regions show the
background estimates obtained  by interpolating the results of the Dalitz plot analyses of the sideband regions.}
\label{fig:fig10}
\end{center}
\end{figure}

We define the \etac signal region as the range 2.910-3.030~\gevcc, which contains 6710 events with (55.2 $\pm$ 0.6)\% purity. 
Sideband regions are defined as the ranges 2.720-2.840~\gevcc \ and 3.100-3.220~\gevcc, respectively.
Figure~\ref{fig:fig9} shows the  Dalitz plot for the \etac signal region, and Fig.~\ref{fig:fig10} shows the corresponding Dalitz plot projections. The Dalitz plot and the mass projections are very similar to the distributions in Ref.~\cite{etac_babar} for the decay $\etac \to \Ks K^{\pm} \pi^{\mp}$.
 
We observe an enhancement in the low mass region of the $\Kp \Km$ mass spectrum due to the presence of the $a_0(980)$, $a_2(1320)$, and $a_0(1450)$ resonances. 
The $K^\pm \piz$ mass spectrum is dominated
by the $K^*_0(1430)$ resonance. We also observe $K^*(892)$ signals in the $K^\pm \piz$ mass spectrum in both the signal and sideband regions. 
We fit the \etac sidebands using an incoherent sum of amplitudes, which includes contributions from the $a_2(1320)$, $K^*(892)$, $K^*_0(1430)$, $K^*_2(1430)$, $K^*(1680)$, and $K^*_0(1950)$ resonances and from an incoherent background. As for the Dalitz plot analysis described in Sec. VII.A, the resulting amplitude fractions are interpolated 
into the \etac signal region and normalized using the results from the fit to the $\Kp \Km \piz$ mass spectrum. The estimated background contributions are indicated by the shaded regions in Fig.~\ref{fig:fig10}.

We perform a Dalitz plot analysis of \etacpiz\ using a procedure similar to that described for the \etaceta  analysis in Sec. VII.A.  
We note that in this case, the amplitude contributions to the $\Kp \Km$ system must have isospin one in order to satisfy isospin
conservation in \etac decay. As discussed in Sec. VII.A, the $K^* \bar K$ amplitudes, again denoted as $K^{*+}K^-$, must be symmetrized in order to conserve C-parity.
We take the $K^*_0(1430)^+\Km$ amplitude as the reference, and so set its phase to zero. The $a_0(980)$ resonance is parameterized as a coupled-channel Breit-Wigner resonance whose parameters
are taken from Ref.~\cite{cbar}.
We do not include an additional $\mathcal{S}$-wave isobar amplitude in the nominal fit.  If we include a $K^{*+}_0(800) \Km$ amplitude, as for example in Ref.~\cite{kappa}, we find that its contribution is consistent with zero.

Table~\ref{tab:tab_dkkpi0} summarizes the amplitude fractions and phases obtained from the fit. Using a method similar  to that described in Sec. VII.C, we divide the Dalitz plot into a number of cells such that the number of expected events in each cell is at least eight. In this case there are 12 free parameters and we obtain $\chi^2/\nu=212/130$. 
We observe a relatively large $\chi^2$ contribution ($\chi^2=19$ for 2 cells) in the lower left corner of the Dalitz plot, where the momentum of the \piz\ is very small;
this may be due to a residual contamination from $\gamma \gamma \to \Kp \Km$ events. 
\begin{table}[ht]
\caption{Results of the Dalitz plot analysis of the $\eta_c \to  \Kp \Km \piz$ channel.}
\label{tab:tab_dkkpi0}
\begin{center}
\vskip -0.2cm
{\small
\begin{tabular}{lr@{}c@{}lr@{}c@{}l}
\hline
 \noalign{\vskip2pt}
Final state & \multicolumn{3}{c}{Fraction \%} & \multicolumn{3}{c}{Phase (radians)}\cr
\hline
\hline
 \noalign{\vskip2pt}
$K^*_0(1430)^+ \Km$ & 33.8 $\pm$ & \, 1.9  $\pm$ & \, 0.4  & 0. \cr
$K^*_0(1950)^+ \Km$ & 6.7 $\pm$  & \, 1.0 $\pm$  & \, 0.3 & -0.67 $\pm$  & \, 0.07 $\pm$  & \, 0.03 \cr
$a_0(980) \piz$ & 1.9 $\pm$  & \, 0.1 $\pm$  & \, 0.2 & 0.38 $\pm$  & \, 0.24 $\pm$  & \, 0.02\cr
$a_0(1450) \piz$ & 10.0 $\pm$ & \, 2.4  $\pm$ & \, 0.8 & -2.4 $\pm$  & \, 0.05 $\pm$  & \, 0.03 \cr
$a_2(1320) \piz$ & 2.1 $\pm$  & \, 0.1 $\pm$  & \, 0.2 &  0.77 $\pm$  & \, 0.20 $\pm$ & \, 0.04\cr
$K^*_2(1430)^+ \Km$ & 6.8 $\pm$  & \, 1.4 $\pm$  & \, 0.3 & -1.67 $\pm$  & \, 0.07 $\pm$  & \, 0.03 \cr
{\it NR} & 24.4 $\pm$  & \, 2.5 $\pm$  & \, 0.6  & 1.49  $\pm$  & \, 0.07 $\pm$  & \, 0.03 \cr
 \noalign{\vskip1pt}
\hline
 \noalign{\vskip1pt}
Sum & 85.8 $\pm$ & \,3.6  $\pm$ & \, 1.2 & \cr
$\chi^2/\nu$ & & \, 212/130  & \, & \cr
 \noalign{\vskip1pt}
\hline
\end{tabular}
}
\end{center}
\end{table}

\begin{figure*}[h]
\begin{center}
\includegraphics[width=16cm]{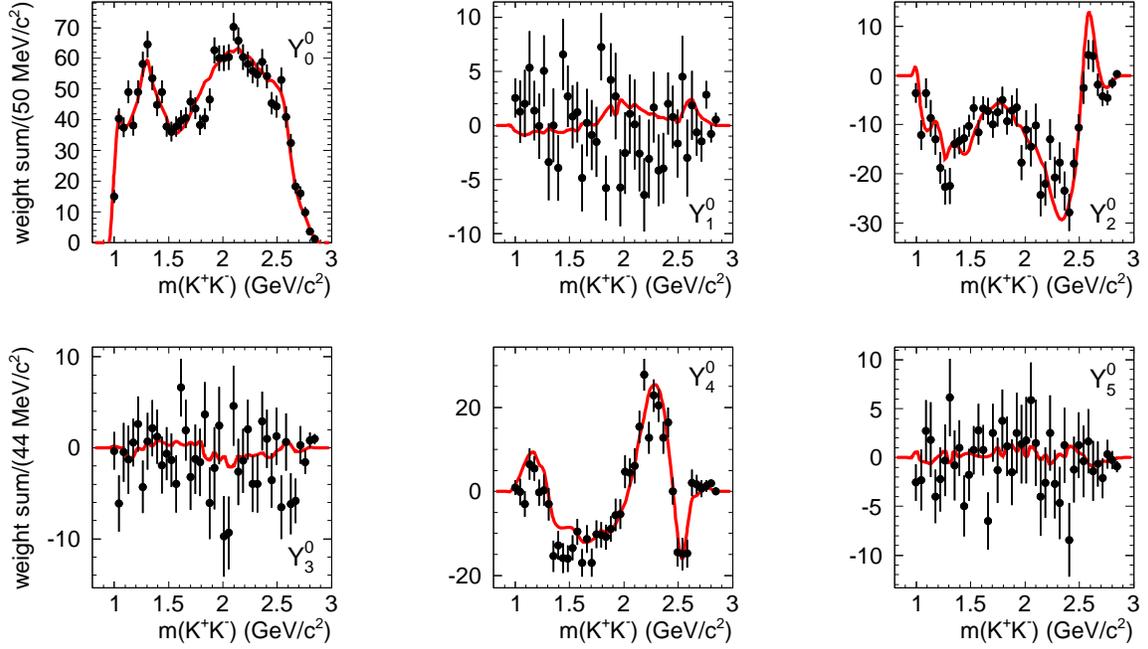}
\caption{Legendre polynomial moments for \etacpiz as a function of $\Kp \Km$ mass. The superimposed curves result from the Dalitz plot analysis described in the text.}
\label{fig:fig11}
\end{center}
\end{figure*}

\begin{figure*}[h]
\begin{center}
\includegraphics[width=16cm]{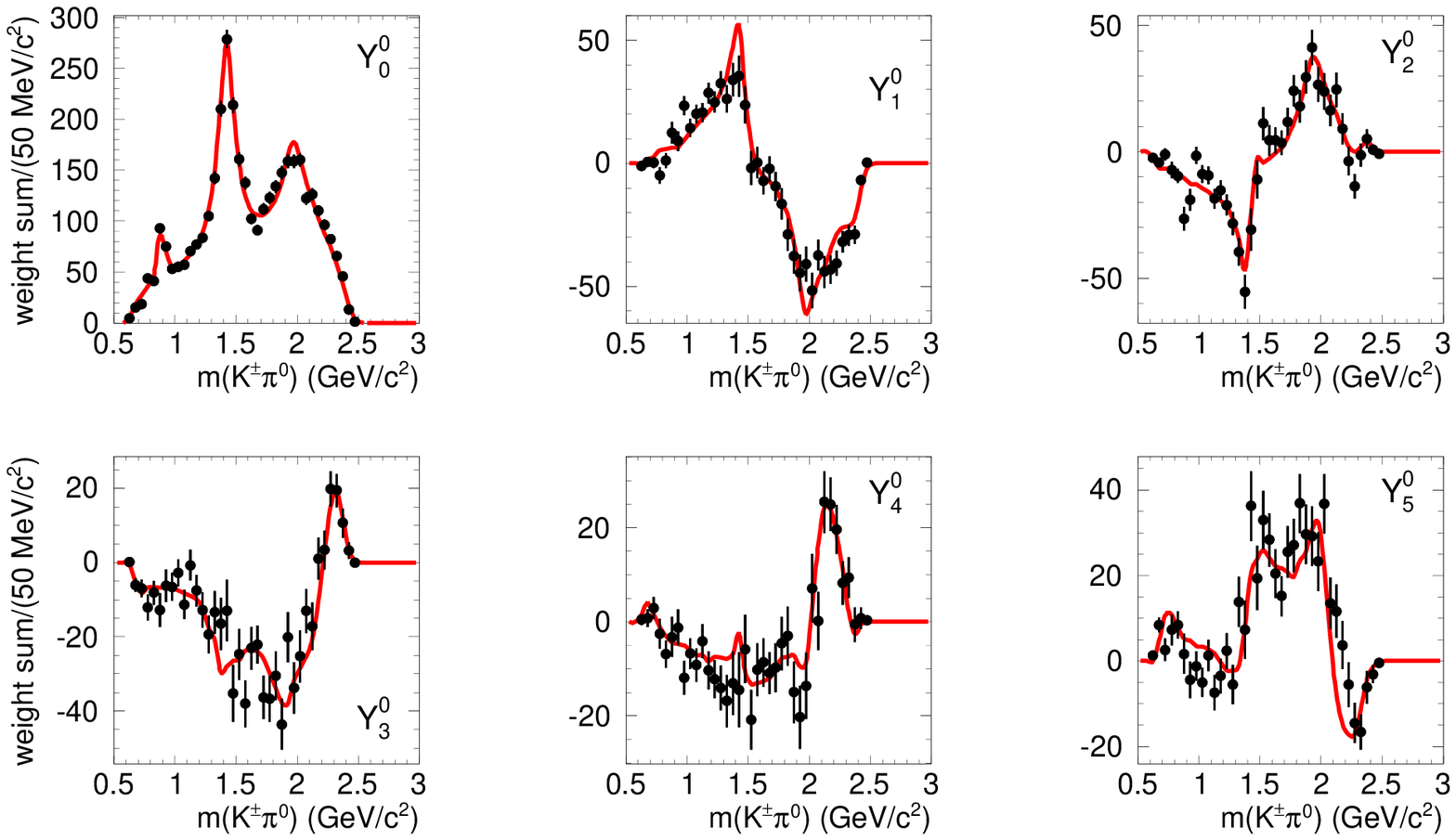}
\caption{Legendre polynomial moments for \etacpiz as a function of $\Kpm \piz$ mass. The superimposed curves result from the Dalitz plot analysis described in the text. The corresponding $\Kp \piz$ and $\Km \piz$ distributions are combined.}
\label{fig:fig12}
\end{center}
\end{figure*}

The Dalitz plot analysis shows a dominance of scalar meson amplitudes with small contributions from spin-two resonances. The $K^*(892)$ contribution is consistent with originating entirely from background. Other spin-one $K^*$ resonances have been included in the fit, 
but their contributions have been found to be
consistent with zero. We note the presence of a sizeable non-resonant contribution. However, in this case the sum of the fractions is significantly lower than 100\%, indicating important interference effects. 
Figure~\ref{fig:fig10} shows the fit projections superimposed on the data, and good agreement is apparent
for all projections.
We compute the uncorrected Legendre polynomial moments $\langle Y^0_L \rangle$ in each  $\Kp \Km$ and $\Kpm \piz$  mass interval by weighting each event by the relevant $Y^0_L(\cos \theta)$ function. These distributions are shown in Figs.~\ref{fig:fig11} and~\ref{fig:fig12}. We also compute the expected Legendre polynomial moments from weighted MC events and compare them with the experimental distributions. We observe satisfactory agreement in all distributions, but we note
that there are regions in which the detailed behavior of some moments is not well reproduced by the fit. This is reflected by the high value of the $\chi^2$ obtained. We have been unable to find additional amplitudes that improve the fit model. This may indicate, for example, that interference between signal and
background is relevant to the Dalitz plot description.

Systematic uncertainty estimates on the fractions and relative phases are obtained by procedures similar to those described in Sec. VII.B. 

\subsection{Determination of the {\boldmath$\protect K^{*}_0(1430)$} parameter values}

\begin{figure}[h]
\begin{center}
\includegraphics[width=8cm]{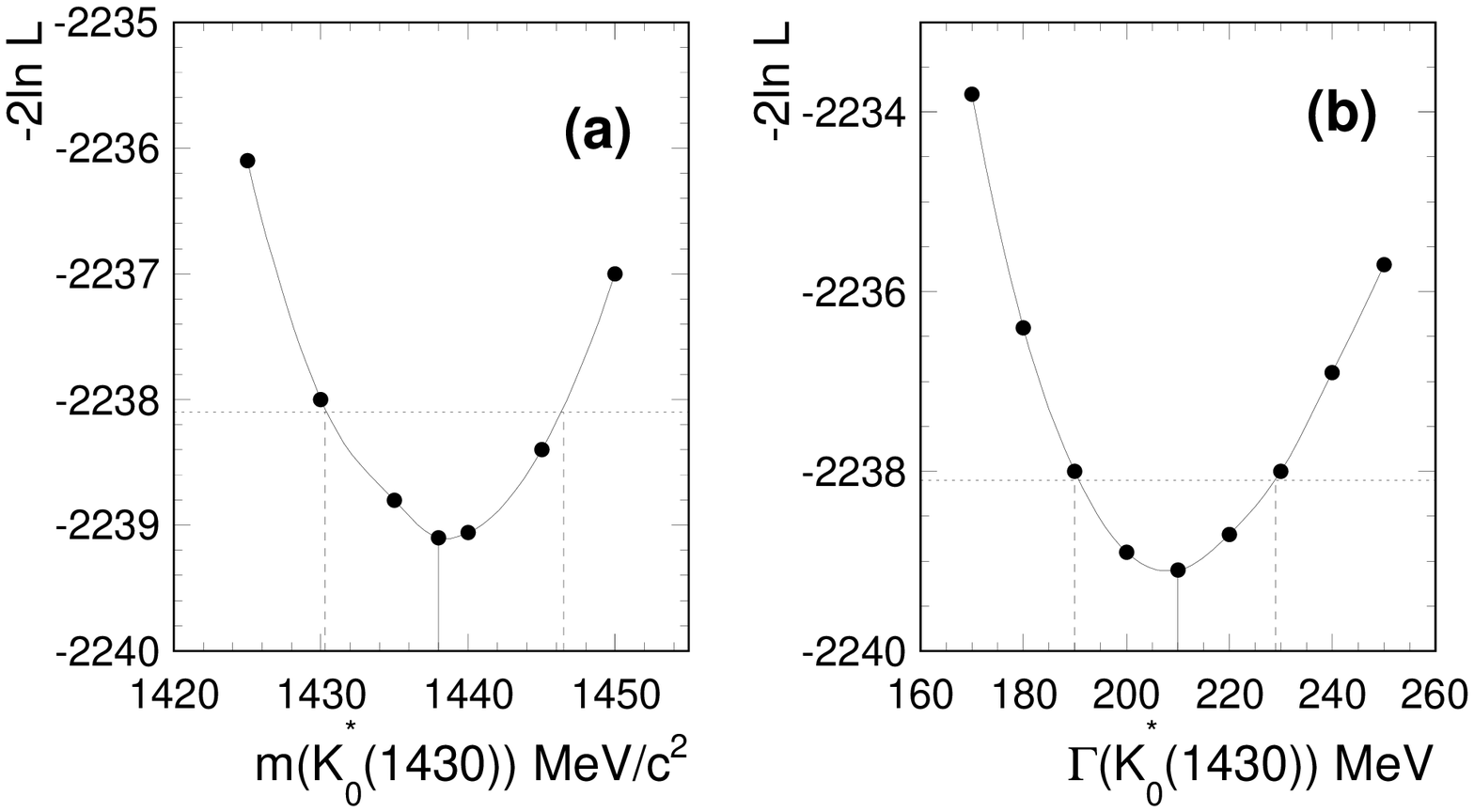}
\caption{Likelihood scans as functions of the $K^{*}_0(1430)$ (a) mass and (b) width. The horizontal (dotted) lines indicate the $\Delta(-2\ln\mathcal{L})=1$ positions, the solid line the likelihood minimum, and the vertical (dashed) lines the uncertainty ranges.}
\label{fig:fig_scan}
\end{center}
\end{figure}

In the Dalitz plot analyses of  $\eta_c \to  \Kp \Km \eta$ and $\eta_c \to  \Kp \Km \piz$, we perform a likelihood scan to obtain the best-fit parameters for the $K^{*}_0(1430)$. 
We use this approach because, in the presence of several interfering scalar-meson resonances, allowing the parameters of the $K^{*}_0(1430)$ to be free results in fit instabilities. The best measurements of the $K^{*}_0(1430)$ parameters have been obtained by the LASS experiment~\cite{lass_kpi}, in which the mass value
$m=1435~\pm~5~\mevcc$ and width value $\Gamma= 279 \pm 6~\mev$ were found for the $K^{*}_0(1430)$~\cite{babar_z}.
First, we fix the mass to 1435~\mevcc\ and examine $-2\ln\mathcal{L}$ as a function of the $K^{*}_0(1430)$ width.
We find that the function has a minimum at 210 \mev for both $\eta_c$ decay modes. We determine the uncertainty by requiring  $\Delta(-2\ln\mathcal{L})=1$. We obtain
$\Gamma=210 \pm 20~\mev$ and $\Gamma=240^{+60}_{-50}~\mev$ from the $\eta_c \to  \Kp \Km \piz$ and $\eta_c \to  \Kp \Km \eta$ scans, respectively.
Fixing the width to 210 \mev, we then scan the likelihood for the $K^{*}_0(1430)$ mass and obtain $m=1438 \pm 8~\mevcc$ for the $\eta_c \to  \Kp \Km \piz$ decay mode. Figure~\ref{fig:fig_scan} shows the results of the likelihood scans.
For the $\eta_c \to  \Kp \Km \eta$ mode, we obtain a minimum at 1435 \mev, but the limited size of the event sample does not permit a useful evaluation of the uncertainty. 
We evaluate systematic uncertainties for the $K^{*}_0(1430)$ parameters by repeating the $\eta_c \to  \Kp \Km \piz$ scans for different values of parameters in the ranges of their statistical uncertainties obtaining
\begin{equation}
\begin{split}
m(K^{*}_0(1430))=1438 \pm 8 \pm 4 \ \mevcc \\
\Gamma(K^{*}_0(1430)) = 210 \pm 20 \pm 12 \ \mev.
\end{split}
\end{equation}
The mass value agrees well with that from the LASS experiment, but the width is approximately three standard deviations smaller than the LASS
result.

\section{{\boldmath$\protect K\eta/K\pi$} Branching ratio for the {\boldmath$\protect K^*_0(1430)$}}

The observation of the $K^*_0(1430)$ in the $K \eta$ and $K \piz$ decay modes permits a measurement of the corresponding branching
ratio.
Taking into account the systematic uncertainty on the fractions of contributing amplitudes, the Dalitz plot analysis of $\eta_c \to K^+ K^- \eta$ decay gives a total $K^*_0(1430)^+\Km$ contribution of
\begin{equation} 
f_{\eta K} = 0.164 \pm 0.042 \pm 0.010.
\end{equation}
Similarly, the Dalitz plot analysis of the $\eta_c \to  K^+ K^-\piz$ decay mode gives a total $K^*_0(1430)^+\Km$ contribution of
\begin{equation}
f_{\piz K} = 0.338 \pm 0.019 \pm 0.004.
\end{equation} 
Using the measurement of $\calR(\etac)$ from Eq. (6), we obtain the $K^*_0(1430)$ branching ratio
\begin{equation} 
\frac{\BR(K^*_0(1430) \to \eta K)}{\BR(K^*_0(1430) \to \pi K)} = \calR(\etac) \frac{f_{\eta K}}{f_{\pi K}} = 0.092 \pm 0.025 \pm 0.010,
\end{equation} 
where $f_{\pi K}$ denotes $f_{\piz K}$ after correcting for the $K^0 \pi$ decay mode.

We note, however, that in the Dalitz plot analyses the amplitude labelled ``{\it NR}'' may be considered to represent an $\mathcal{S}$-wave
$K \pi$ or $K \eta$ system in an orbital $\mathcal{S}$-wave state with respect to the bachelor kaon. As such, the {\it NR} amplitude  has structure
similar to that of the $K^*_0(1430)^+ \Km$ amplitudes, and hence may influence the associated fractional intensity contributions 
through interference effects. Therefore, we assess an additional systematic uncertainty on the value of the branching ratio given in Eq. (13); this is done in order to account for the impact of the {\it ad hoc} nature of the representation of the {\it NR} amplitude.

For example, if we denote the relative phase between the {\it NR} and $K^*_0(1430)^+ \Km$  amplitudes by $\phi_{NR}$, the value listed
in Table~\ref{tab:tab_dkkpi0} is approximately $+\pi/2$, so that the interference term between the amplitudes behaves like the imaginary part
of the $K^*_0(1430)$ BW amplitude. This has the same mass dependence as the squared modulus of the BW, and it follows that
the interference term causes the fractional contribution associated with the $K^*_0(1430)^+ \Km$ amplitude to be reduced.

We study the correlation between $\phi_{NR}$ and the $K^*_0(1430)^+ \Km$ fraction $f_{\piz K}$ by 
performing different fits in which $\phi_{NR}$ is arbitrarily fixed to different values from 0 to $3 \pi/2$. We observe a correlation between $f_{\piz K}$ and $\phi_{NR}$ with $f_{\piz K}$ varying from $(33.3 \pm 1.8)\%$ at $\phi_{NR}=\pi/2$ to $(67.0\pm2.2)\%$ at $\phi_{NR}=3\pi/2$.
To estimate the systematic uncertainty related to this effect, we remove the non-resonant contribution in both the $\etac \to \Kp \Km \eta$ and $\etac \to \Kp \Km \piz$ Dalitz plot analyses. We obtain changes of the negative log likelihood $\Delta(-2{\ln} \mathcal{L})$=+319
and $\Delta(-2{\ln} \mathcal{L})$=+20 for $\etac \to \Kp \Km \piz$ and $\etac \to \Kp \Km \eta$ decays, respectively, for the reduction by 2 parameters. The corresponding variation of the $f_{\eta K}/f_{\piz K}$ fraction is -0.023 and we assign this as the associated systematic uncertainty.
We thus obtain
\begin{equation} 
\frac{\BR(K^*_0(1430) \to \eta K)}{\BR(K^*_0(1430) \to \pi K)} = \calR(\etac) \frac{f_{\eta K}}{f_{\pi K}} = 0.092 \pm 0.025^{+0.010}_{-0.025}.
\end{equation}

The LASS experiment studied the reaction $K^- p \to K^- \eta p$ at 11 \gevc~\cite{lass}. The $K^- \eta$ mass spectrum is dominated by the presence of the
$K^*_3(1780)$ resonance with no evidence for $K^*_0(1430) \to K \eta$ decay.
However, from Ref.~\cite{lass_kpi}
\begin{equation}
\Gamma(K^*_0(1430)\to K \pi)/\Gamma(K^*_0(1430))=0.93 \pm 0.04 \pm 0.09,
\end{equation}
which is not in conflict with the presence of a small branching fraction for the $K \eta$ decay mode.

\section{Implications of the {\boldmath$\protect K^*_0(1430)$} branching ratio for the pseudoscalar meson mixing angle}

  As noted in Sec. VIII, there is no evidence for $K^*_0(1430)$ production in
the reaction $K^- p \to K^- \eta p$ at 11 \gevc~\cite{lass}. There is also no evidence
for $K^*_2(1430)$ production in this reaction, and a 0.92\% upper limit on the
branching ratio $\BR(K^*_2(1430) \to K \eta)/\BR(K^*_2(1430) \to K \pi)$ is obtained at
95\% confidence level. In Ref.~\cite{lass}, this small value is understood in the
context of an SU(3) model with octet-singlet mixing of the $\eta$ and $\eta'$~\cite{lipkin}.
For even angular momentum $l$ (i.e., D-type coupling), it can be shown~\cite{nag_the}
that a consequence of the resulting $K^* \bar K \eta$ couplings is
\begin{equation}
\begin{split}
R_l = & \frac{\BR(K^*_l \to K \eta)}{\BR(K^*_l \to K \pi)} \\ 
 = &\frac{1}{9}(\cos \theta_p + 2\cdot\sqrt{2}\cdot \sin \theta_p)^2 \cdot (q_{K \eta}/q_{K \pi})^{2l+1}
\end{split}
\end{equation}
where $q_{K \eta}$ ($q_{K \pi}$) is the kaon momentum in the $K \eta$ ($K \pi$) rest
frame at the $K^*$ mass and $\theta_p$ is the SU(3) singlet-octet mixing angle
for the pseudoscalar meson nonet. We note that $R_l$ equals zero if
$\tan \theta_p = - [1/(2 \cdot \sqrt{2})]$ (i.e., $\theta_p = - 19.7^{\circ}$).

For $l=2$, the upper limit $R_2 = 0.0092$ corresponds to $\theta_p = - 9.0^{\circ}$
and the central value yields $\theta_p = - 11.4^{\circ}$.

In the present analysis, we obtain the value $R_0=0.092^{+0.027}_{-0.035}$,
where we have combined the statistical and systematic uncertainties in
quadrature. The corresponding value of $\theta_p$ is
$(3.1^{+3.3}_{-5.0})^{\circ}$, which differs by about 2.9 standard deviations from the result obtained from 
the $K^*_2(1430)$ branching ratio.

The value of $R_2$ from Ref.~\cite{lass} is in reasonable agreement with the 
analysis reported
in Ref.~\cite{gilman}, which concludes that $\theta_p \sim - 20^{\circ}$ is consistent with 
experimental evidence from many different sources, although $\theta_p \sim - 10^{\circ}$ 
cannot be completely ruled out. In addition, a lattice QCD calculation~\cite{UKQCD} 
yields $\theta_p = (-14.1 \pm 2.8)^{\circ}$ for the value of the octet-singlet 
mixing angle, in good agreement with the spin-two result and the conclusion of
Ref.~\cite{gilman}, but differing by about three standard deviations from the spin-zero measurement. 
However, in Ref.~\cite{feldmann} it is argued that it is necessary to consider separate
octet and singlet mixing angles for the pseudoscalar mesons. For the octet, 
experimental data from many sources indicate a mixing angle of $\sim - 20^{\circ}$, 
whereas for the singlet the values are almost entirely in the range from 
zero to $-10^{\circ}$. The analysis of Ref.~\cite{feldmann} may be able to provide an 
explanation for the small value of the magnitude of $\theta_p$ extracted from 
our measurement of the $K^*_0(1430)$ branching ratio by using the model
suggested in Ref.~\cite{lipkin}.

\section{Summary}

We have studied the processes $\gg\to \kketa$ and $\gg\to \kkpiz$ using a data sample corresponding to an integrated luminosity 
of 519~\invfb\ recorded with the \babar\ detector at the SLAC PEP-II
asymmetric-energy \epem\ collider at center-of-mass energies at and near the
$\Upsilon(nS)$ ($n = 2,3,4$) resonances.
We observe \etacpiz\ decay and obtain the first observation of \etaceta\ decay, measure their relative branching fractions, and perform a Dalitz plot 
analysis for each decay mode. 
The Dalitz plot analyses demonstrate the dominance of quasi-two-body amplitudes involving scalar-meson resonances. 
In particular, we observe significant branching fractions for $\etac \to f_0(1500) \eta$ and $\etac \to f_0(1710) \eta$. Under the hypothesis of a gluonium content
in these resonances, similar decay branching fractions to $\pi \pi$ and $K \bar K$ are expected. To obtain these measurements, it would be useful to study 
$\etac \to \eta \pi \pi$, $\etac \to \eta' \Kp \Km$, and $\etac \to \eta' \pip \pim$ decays.
We obtain the first observation of $\Kstarz \to K \eta$ decay, and
measure its branching fraction relative to the $K \pi$ mode to be $\calR(K^*_0(1430)) = \frac{\BR(K^*_0(1430) \to K \eta)}{\BR(K^*_0(1430) \to K \pi)} = 0.092 \pm 0.025^{+0.010}_{-0.025}$.
This observation is not in complete agreement with the SU(3) expectation that the $K \eta$ system 
almost decouple from 
even-spin $K^*$ resonances~\cite{lass}. Based on the Dalitz plot analysis of  $\eta_c \to  \Kp \Km \piz$, we measure the $K^{*}_0(1430)$ parameters and obtain $m=1438 \pm 8 \pm 4\ \mevcc$ and $\Gamma=210 \pm 20 \pm 12 \ \mev$. 
We observe evidence for \etactwopiz\ decay, first evidence for \etactwoeta\ decay, and measure their relative branching fraction.

\section{Acknowledgements}
We are grateful for the 
extraordinary contributions of our \pep2\ colleagues in
achieving the excellent luminosity and machine conditions
that have made this work possible.
The success of this project also relies critically on the 
expertise and dedication of the computing organizations that 
support \babar.
The collaborating institutions wish to thank 
SLAC for its support and the kind hospitality extended to them. 
This work is supported by the
US Department of Energy
and National Science Foundation, the
Natural Sciences and Engineering Research Council (Canada),
the Commissariat \`a l'Energie Atomique and
Institut National de Physique Nucl\'eaire et de Physique des Particules
(France), the
Bundesministerium f\"ur Bildung und Forschung and
Deutsche Forschungsgemeinschaft
(Germany), the
Istituto Nazionale di Fisica Nucleare (Italy),
the Foundation for Fundamental Research on Matter (The Netherlands),
the Research Council of Norway, the
Ministry of Education and Science of the Russian Federation,
Ministerio de Economia y Competitividad (Spain), and the
Science and Technology Facilities Council (United Kingdom).
Individuals have received support from 
the Marie-Curie IEF program (European Union), the A. P. Sloan Foundation (USA) 
and the Binational Science Foundation (USA-Israel).

\renewcommand{\baselinestretch}{1}

\end{document}